\documentclass[twocolumn, showpacs,preprintnumbers,amsmath,amssymb]{revtex4}

\usepackage{amsmath,amssymb}
\usepackage{bm}
\usepackage{graphicx}
\usepackage{dcolumn}
\usepackage{bm}
\usepackage{color}

\begin{document}

\title{Maxwell stress in fluid mixtures}
\author{Takahiro Sakaue$^{1,2}$
\footnote{E-mail: sakaue@phys.kyushu-u.ac.jp}
 and Takao Ohta$^{3}$}
\affiliation{$^1$Department of Physics, Kyushu University 33, Fukuoka, 812-8581, Japan \\
$^2$PREST, JST, 4-1-8 Honcho Kawaguchi, Saitama 332-0012, Japan \\
$^3$Department of Physics, Kyoto University, Kyoto, 606-8502, Japan 
}
\date{\today}


\begin{abstract}
We examine the structure of Maxwell stress in binary fluid mixtures under an external electric field and discuss its consequence. In particular, we show that, in immiscible blends, it is intimately related to the statistics of domain structure. This leads to a compact formula, which may be useful in the investigation of electro-rheological effects in such systems. 
The stress tensor calculated in a phase separated fluid under a steady electric field is in a good agreement with recent experiments.

\end{abstract}


\maketitle

{\it Introduction}---
Like any other complex fluid, blends of immiscible fluids exhibit rich phase behaviors and dynamics under external fields~\cite{Tucker}. Their domain interface is inherently soft so that the configuration is easily deformed, from which a large stress contribution shows up. Rheological properties under flow fields have been well understood at a semi-quantitative level~\cite{Doi_Ohta91} 
where the so-called interfacial tensor plays a central role to capture the statistical properties of complicated interconnected domain structures. 
%

The shape change of droplets and interfacial instabilities under an electric field in immiscible fluids having different dielectric constants have been studied both experimentally and theoretically for many years~\cite{Carson, Melcher, Kotaka, Krause}. Structural transitions induced by electric field have also been investigated in microphase separation in block copolymers~\cite{Amundson, Fukuda, Tsori}. However, these previous studies were concerned mainly with the morphological change of domains. It should be emphasized that a change of domain structures drastically affects flow behavior of the system and, therefore, produces a unique rheological effect. This type of electro-rheology is of fundamental importance since the cross coupling between flow field and electric field is relevant to characterizing the departure from equilibrium. Recently, experiments of electro-rheology in phase separating fluids have been conducted ~\cite{Orihara97, Pan97, Koyama98, Ha00, Orihara01}. A key physical quantity is the electric (Maxwell) stress, which is expected to be intimately related with the spatial domain structures.

%

In this paper, we discuss a fundamental relation between the Maxwell stress and domain configurations in fluid mixtures.
We  start with the basic equations for phase separation dynamics in which the free energy functional contains  the electro-static energy. First, we show that  the reversible mode coupling term in the dynamic equation for the local velocity produces exactly the Maxwell stress. We then eliminate the local electric field ${\vec E}({\vec r})$ and obtain an expression to the Maxwell stress in terms of the external electric field ${\vec E}^{ex}$, which involves a non-local coupling among concentration fluctuations of the induced dipole type.
Then, for immiscible blends, the use of the basic statistical property of random configuration of interfaces at short distance (known as the Porod law) makes us propose that the macroscopic Maxwell stress contribution can be represented in terms of the interfacial tensor. This provides us with a useful formula which connects  the Maxwell stress with arbitrary configuration of domain structures.

{\it Basic equations}---
We consider a binary fluid of A and B components, the local volume fractions of which are represented by $\phi_A({\vec r})$ and $\phi_B({\vec r})$.
The free energy for $\phi({\vec r})=\phi_A({\vec r})-\phi_B({\vec r})$ consists of two parts:
\begin{eqnarray}
F\{\phi ({\vec r})\} =F_1\{\phi({\vec r})\}+F_2 \{\phi({\vec r}), \vec{E}({\vec r})\},
\label{Free}
\end{eqnarray}
where
\begin{eqnarray}
F_1\{\phi({\vec r})\} = \int d\vec{r} \left[ \frac{K}{2} ({\vec \nabla}\phi({\vec r}))^2 +f(\phi({\vec r})) \right]
\label{F1}
\end{eqnarray}
and
\begin{eqnarray}
F_2 \{\phi({\vec r}), \vec{E}\} = -\frac{1}{8\pi}\int d\vec{r} \  \epsilon[\phi({\vec r})]\vec{E}(\vec{r} )\cdot\vec{E}(\vec{r} )
\label{F2}
\end{eqnarray}
The constant $K$ is positive, $f(\phi) $ is a polynomial of $\phi$ with two minima. The second free energy $F_2$ is the electric energy where $\vec{E}(\vec{r} )$ is an electric field. The dielectric constant $\epsilon$ is assumed to depend on $\phi$  as $\epsilon = {\bar \epsilon} + \delta \epsilon   \phi (\vec{r} )$ with ${\bar \epsilon} = (\epsilon_A + \epsilon_B)/2$ and $\delta \epsilon = (\epsilon_A - \epsilon_B)/2$ where the dielectric constant of A (B) compound is denoted as $\epsilon_A$ ($\epsilon_B$).

Macrophase separation is governed by the following set of equations for $\phi(\vec{r} )$ and the velocity field $\vec{v}(\vec{r} )$. The local volume fraction $\phi$ obeys
\begin{eqnarray}
\frac{\partial\phi}{\partial t} + {\vec \nabla}\cdot(\vec{v}\phi) = {\vec \nabla} \cdot L {\vec \nabla} \left(\frac{\delta F}{\delta\phi} \right)
\label{phi}
\end{eqnarray}
where $L$ is a mobility coefficient. We introduce the potential $U$ as $\vec{E}(\vec{r} )=\vec{\nabla}U$. The functional derivative of $F_2$ with respect to $U$ 
\begin{eqnarray}
\frac{\delta F_2 \{\phi({\vec r}), \vec{E}\}}{\delta U} =0
\label{FV}
\end{eqnarray}
gives us the Maxwell equation
\begin{eqnarray}
\vec{\nabla}(\epsilon  \vec{E} )=\nabla_{\beta}(\epsilon  \nabla_{\beta}U )=0
\label{E2} 
\end{eqnarray}
(the repeated indices imply summation throughout the paper.)
The condition 
\begin{eqnarray}
\vec{\nabla} \times  \vec{E} = 0
\label{E1} 
\end{eqnarray}
is automatically satisfied.

The local velocity field is governed by
\begin{eqnarray}
\rho \frac{\partial\vec{v}}{\partial t} + \rho (\vec{v}\cdot{\vec \nabla})\vec{v} = -{\vec \nabla} p 
-\phi \vec{\nabla}\frac{\delta F}{\delta \phi}+\eta_0  {\nabla}^2\vec{v}
\label{v}
\end{eqnarray}
where $\rho$ is the fluid density, $p$ is determined to satisfy the incompressibility condition ${\vec \nabla} \cdot \vec{v}=0$, and the viscosity $\eta_0$ is assumed to be constant and independent of $\phi ({\vec r})$.  Equation~(\ref{v}) may contain another term like $U\vec{\nabla}(\delta F_2/\delta U)$. However, this term vanishes because of the relation (\ref{FV}).

It is well known that $-\phi \vec{\nabla}(\delta F_1/\delta \phi)$ is related with the stress tensor $\sigma^D({\vec r})$ arising from the gradient term 
in (\ref{F1})
as
~\cite{Batchelor70,Onuki_text}
\begin{eqnarray}
-\phi  ({\vec r}) \nabla_{\alpha}\frac{\delta F_1}{\delta \phi}=\nabla_{\beta}\sigma^D_{\alpha\beta}({\vec r})
\label{stressD1}
\end{eqnarray}
The off-diagonal parts of $\sigma^D({\vec r})$ are given by
\begin{eqnarray}
\sigma^D_{\alpha\beta}({\vec r})=-K(\nabla_{\alpha}\phi({\vec r}))( \nabla_{\beta}\phi({\vec r}))
\label{stressD2}
\end{eqnarray}
In parallel, 
we may write  the electric contribution from $F_2$ as
\begin{eqnarray}
-\phi ({\vec r})\nabla_{\alpha}\frac{\delta F_2}{\delta \phi}=\nabla_{\beta}\sigma^M_{\alpha\beta} ({\vec r})
\label{stressM0}
\end{eqnarray}
where $\sigma^M_{\alpha \beta} ({\vec r})$ is 
identified with 
the local Maxwell stress defined by  \cite{Landau}
\begin{eqnarray}
\sigma^M_{\alpha\beta} ({\vec r})=\frac{\epsilon({\vec r})}{4\pi} E_{\alpha}({\vec r})E_{\beta}({\vec r})
\label{stressM}
\end{eqnarray}
To prove this, let us rewrite the left-hand side of eq.~(\ref{stressM0}) as
\begin{eqnarray}
-\phi \nabla_{\alpha}\left(\frac{\delta F_2}{\delta \phi}\right)&=&+(\nabla_{\alpha}\phi) \frac{\delta F_2}{\delta \phi} \nonumber \\
&=&-\frac{1}{8\pi}(\nabla_{\alpha} \epsilon)(\nabla_{\beta}U)(\nabla_{\beta}U) \nonumber \\
&=&\frac{1}{4\pi} \epsilon(\nabla_{\alpha}\nabla_{\beta}U)(\nabla_{\beta}U)
\label{tensor1}
\end{eqnarray}
where we have 
absorbed a term in the form of ${\vec \nabla} X$ into the term ${\vec \nabla} p$ in eq. (\ref{v}) 
to reach the final expression.
Then, the relation (\ref{stressM0}) follows by calculating the right-hand side with the definition of the Maxwell stress (eq.~(\ref{stressM}))
\begin{eqnarray}
\nabla_{\beta} \sigma^{M}_{\alpha \beta}&=&\frac{1}{4\pi} \nabla_{\beta}[\epsilon(\nabla_{\alpha}U)(\nabla_{\beta}U)] \nonumber \\
&=&\frac{1}{4\pi} \epsilon(\nabla_{\beta}U)(\nabla_{\alpha}\nabla_{\beta}U),
\label{tensor3}
\end{eqnarray}
where we have used the relation (\ref{E2}).

{\it  Macroscopic stress}---
In principle, the dynamics and rheology of the system can be studied by the above set of basic equations. 
However, this usually requires rather intense numerical computations, and a more coarse grained description is called for both to develop the analytically tractable theory and to get a deeper physical insight.
In this direction, Doi and Ohta proposed a semi-phenomenological rheological constitutive equation for the immiscible blends under flow~\cite{Doi_Ohta91}, where it was important to realize the stress expression in terms of the domain configurations, i.e., the stress $\sigma^{D}_{\alpha \beta}=\langle \sigma^{D}_{\alpha \beta}({\vec r})\rangle $ arising from the gradient term (eq.~(\ref{stressD2})) can be written as~\cite{Batchelor70,Onuki_text}
\begin{eqnarray}
\sigma^{D}_{\alpha \beta}= \langle \sigma^{D}_{\alpha \beta}({\vec r})\rangle \simeq  -\Gamma q_{\alpha \beta} . \label{sigma_D_domain}
\end{eqnarray}
Here the bracket indicates the averaging over the spatial configurations, $\Gamma \simeq K/\xi$ is the interfacial tension with $\xi$ being the interfacial thickness and the interfacial tensor is defined as
\begin{eqnarray}
q_{\alpha \beta} = \frac{1}{V} \int dS \ (n_{\alpha}n_{\beta})  \label{sigma_q}
\end{eqnarray}
where $\int dS \ ()$ is the surface integral, ${\vec n}$ is the unit vector normal to the interface.
In the electro-rheological problems, it would be therefore desirable to transform the Maxwell stress by treating the $F_2$ term much in the same way as the gradient term in the free energy $F_1$. Below, we shall seek such a meaningful expression for the Maxwell stress in terms of the external (but not local) electric field as well as the domain configurations.

Equations (\ref{E2}) and (\ref{E1}) are solved by a perturbation expansion in terms of 
$\delta \epsilon = (\epsilon_A -\epsilon_B)/2$~\cite{Onuki_Doi92}. The electric field is expanded as
\begin{eqnarray}
\vec{E} =\vec{E}^{(0)}+\delta\epsilon  \vec{E}^{(1)}+O((\delta\epsilon^2))
\label{Eexpand}
\end{eqnarray}
The external field $\vec{E}^{ex}$ constant in space satisfies the zeroth order solution, i.e., $\vec{E}^{(0)}=\vec{E}^{ex}$.  The first order solution should satisfy
\begin{eqnarray}
 \vec{\nabla}( \phi \vec{E}^{ex}) +{\bar \epsilon} \vec{\nabla}\vec{E}^{(1)}= 0
\label{E11} \\
\vec{\nabla} \times  \vec{E}^{(1)} = 0
\label{E21}
\end{eqnarray}
If we put $\vec{E}^{(1)}$ in terms of the Fourier transformation $ E_{\alpha, {\vec k}}^{(1)} = \int d{\vec r} \  E_{\alpha}^{(1)}({\vec r}) e^{i {\vec k}\cdot{\vec r}}$
\begin{eqnarray}
\delta \epsilon E^{(1)}_{\beta, {\vec k}}= - \frac{\delta \epsilon}{{\bar \epsilon}}E_{\alpha}^{ex} G^{\alpha \beta}_{{\vec k}} \phi_{\vec k}
\label{E111}
\end{eqnarray}
this satisfies eqs. (\ref{E11}) and (\ref{E21}), where
\begin{eqnarray}
G^{\alpha \beta}_{{\vec k}}  = \frac{k_{\alpha}k_{\beta}}{k^2}
\end{eqnarray}
and its inverse Fourier transformation is given by
\begin{eqnarray}
G^{\alpha \beta}(\vec{r}, \vec{r'})=G(\vec{r}, \vec{r'})\nabla^{\alpha}\nabla'^{\beta}
\label{G1}
\end{eqnarray}
with
$-\nabla^2G(\vec{r}, \vec{r'})=\delta(\vec{r}-\vec{r'})$.
Using eq.~(\ref{E111}), the Maxwell stress eq. (\ref{stressM}) is written 
up to the order of $(\delta\epsilon)^2$ as
\begin{eqnarray}
\sigma^M_{\alpha\beta}({\vec r})&=&\sigma^{M_0}_{\alpha \beta} \nonumber \\
&-&K_{\epsilon} E^{ex}_{\alpha}E^{ex}_{\delta}\int d\vec{r'}G^{\beta \delta}(\vec{r}, \vec{r'})\phi(\vec{r})\phi(\vec{r'}) \nonumber \\
&-&K_{\epsilon} E^{ex}_{\beta}E^{ex}_{\delta}\int d\vec{r'}G^{\alpha \delta}(\vec{r}, \vec{r'})\phi(\vec{r})\phi(\vec{r'})  \nonumber \\
&+&K_{\epsilon}E^{ex}_{\gamma}E^{ex}_{\delta} \int d\vec{r'}\int d\vec{r''}G^{\alpha \gamma}(\vec{r}, \vec{r'})\nonumber \\
&\times&G^{\beta \delta}(\vec{r}, \vec{r''})\phi(\vec{r'})\phi(\vec{r''}) 
\label{stressM1}
\end{eqnarray}
where $K_{\epsilon}=(\delta \epsilon)^2/(4 \pi {\bar \epsilon})$ and
$\sigma^{M_0}_{\alpha \beta}=({\bar \epsilon}/4\pi)E^{ex}_{\alpha}E^{ex}_{\beta} $
is the trivial average term, and will be omitted in what follows.


After averaging over the system volume $V$, a macroscopic stress $\sigma^{M}_{\alpha \beta} = \langle \sigma^{M}_{\alpha \beta} \rangle = V^{-1}\int_V d{\vec r}\  \sigma^{M}_{\alpha \beta}({\vec r})$ is obtained as
\begin{eqnarray}
\sigma^{M}_{\alpha \beta}=
&-&\frac{K_{\epsilon}}{V} E^{ex}_{\alpha}E^{ex}_{\delta}\int_{\vec {k}}G^{\beta \delta}_{\vec{k}}\langle \phi_{\vec{k}} \  \phi_{-\vec{k}} \rangle 
\nonumber \\
&-&\frac{K_{\epsilon}}{V} E^{ex}_{\beta}E^{ex}_{\delta}\int_{\vec {k}}G^{\alpha \delta}_{\vec{k}} \langle \phi_{\vec{k}} \ \phi_{-\vec{k}} \rangle \nonumber \\
&+&\frac{K_{\epsilon}}{ V} E^{ex}_{\gamma}E^{ex}_{\delta}\int_{\vec {k}}G^{\alpha \gamma}_{\vec{k}}G^{\beta \delta}_{-\vec{k}} \ \langle \phi_{\vec{k}} \ \phi_{-\vec{k}}\rangle \nonumber \\
=&-&\frac{K_{\epsilon}}{V} \int_{{\vec k}}\frac{(k_{\alpha} E_{\beta}^{ex} +k_{\beta} E_{\alpha}^{ex}) ({\vec k}\cdot {\vec E}^{ex})}{k^2} \langle \phi_{{\vec k}} \ \phi_{-{\vec k}} \rangle \nonumber \\
&+& \frac{K_{\epsilon}}{V} \int_{{\vec k}}\frac{k_{\alpha} k_{\beta}({\vec k} \cdot {\vec E}^{ex})^2}{k^4} \langle \phi_{{\vec k}} \ \phi_{-{\vec k}} \rangle
\label{sigma_M_2}
\end{eqnarray}


{\it Domain structures}---
The system governed by eqs. (\ref{phi}) and (\ref{v}) generally undergoes macrophase separation in which domain coarsening proceeds.
There are typically two procedures in order to examine a response from steady domain structures. One is to study the late stage of the phase separation process where droplets make a quick response to the electric field compared with the coarsening dynamics.  The other is to apply steady shear flow so that the size of droplets remains finite due to their break-up and reconnection. 
We shall show below that in such situations, the Maxwell stress can be evaluated in terms of the domain structure characteristics.
To this end, we apply a pre-averaging approximation to the double kernel term in eqs.~(\ref{stressM1}) and~(\ref{sigma_M_2}) as following:
\begin{eqnarray}
& &\frac{K_{\epsilon}}{V} \int_{{\vec k}}\frac{k_{\alpha} k_{\beta}({\vec k} \cdot {\vec E}^{ex})^2}{k^4} \langle \phi_{{\vec k}} \ \phi_{-{\vec k}} \rangle \nonumber \\
\Rightarrow&&
C \frac{K_{\epsilon}}{2V} E_{\delta}^{ex}E_{\gamma}^{ex} \int_{{\vec k}} \frac{k_{\alpha}k_{\gamma}}{k^2} \langle \phi_{{\vec k}} \ \phi_{-{\vec k}} \rangle \left\langle \frac{k_{\beta}k_{\delta}}{k^2} \right\rangle_{p.a} \nonumber \\
&+& (\alpha \leftrightarrow \beta) \nonumber \\
&\simeq& \frac{K_{\epsilon}}{2 V}\int_{{\vec k}}\frac{(k_{\alpha}E_{\beta}^{ex} + k_{\beta}E_{\alpha}^{ex}) ({\vec k}\cdot{\vec E}^{ex})}{k^2} \langle \phi_{{\vec k}} \ \phi_{-{\vec k}} \rangle 
\label{preaverage}
\end{eqnarray}
where the pre-averaging indicates $\langle k_{\alpha}k_{\beta}/k^2\rangle_{p.a} = 1/3 \ \delta_{\alpha \beta}$ and the constant $C=3$ is determined by requiring that the result becomes consistent after taking the trace.
Substituting eq.~(\ref{preaverage}) into eqs.~(\ref{stressM1}) or~(\ref{sigma_M_2}), we obtain
\begin{eqnarray}
 \sigma^{M}_{\alpha \beta}&=&-\frac{K_{\epsilon}}{2}  E^{ex}_{\alpha}E^{ex}_{\delta} \nonumber \\
  &\times& \int d{\vec r} G^{\beta \delta}({\vec r}_1,{\vec r}_1+{\vec r})  \langle\phi({\vec r}_1)\phi({\vec r}_1+{\vec r})\rangle   \nonumber \\
&+& (\alpha \leftrightarrow \beta)\nonumber \\
&=& -\frac{K_{\epsilon}}{2}  E^{ex}_{\alpha}E^{ex}_{\delta}  \int d{\vec r} G({\vec r})g_1^{\beta \delta}({\vec r}) + (\alpha \leftrightarrow \beta) 
\label{sigma_M_preaverage}
\end{eqnarray}
where we 
have
introduced the pair correlation of $\phi$ and ${\vec \nabla} \phi$ as
\begin{eqnarray}
g({\vec r}) &=&  \langle \phi({\vec r}_1)\phi({\vec r}_1+{\vec r}) \rangle  \\
g_1^{\alpha \beta}({\vec r}) &=&   \langle\nabla^{\alpha}_{1}\phi({\vec r}_1) \nabla^{\beta}_{2}\phi({\vec r}_2) \rangle |_{ {\vec r}_2={\vec r}_ 1+{\vec r} }
\end{eqnarray}
Equation~(\ref{sigma_M_preaverage}) can be further transformed with the aid of the {\it Porod law}, which is generally valid in systems with domain structures where the interface boundaries are rather sharp; $R \gg \xi$ with $R$ being the typical domain scale and the order parameter $\phi ({\vec r})$ takes the stable value (normalized as $\phi_0=1$) except for the interface regions~\cite{Porod51}. 
These conditions are satisfied at the late stage of phase separation.

In the length scale $ \xi \ll r \ll R$, the structure factor $V \int d{\vec r} \  g({\vec r}) e^{i {\vec k} \cdot {\vec r}} = \langle \phi_{{\vec k}}\phi_{-{\vec k}} \rangle$ exhibits the Porod tail $\langle \phi_{{\vec k}}\phi_{-{\vec k}} \rangle \simeq Q k^{-4}$, where $Q=q_{\alpha \alpha}=\int dS  /V$ is the interface area density. This can be written in the real space as $g_1^{\alpha \alpha}(r) \simeq 2Q/r$ where we explicitly write the coefficient for the three-dimensional case~\cite{Onuki_text}.

Performing the integration in eq.~(\ref{sigma_M_preaverage}), we arrive at our main result
\begin{eqnarray}
\sigma^{M}_{\alpha \beta}  
&\simeq& - \frac{K_{\epsilon}}{2} E^{ex}_{\alpha}E^{ex}_{\delta}  \int_0^R dr \ 4 \pi r^2 \left( \frac{1}{r} \right) \left(\frac{2 q_{\delta \beta}}{r} \right) \nonumber \\
&+&  (\alpha \leftrightarrow \beta)  \nonumber \\
&=& - \Gamma  (   q_{\alpha \delta}  \ s_{\delta \beta} +  q_{\beta \delta}  \ s_{\delta \alpha} )
\label{M_stress_expression}
\end{eqnarray}
where we cut off the integral at the domain scale, and the contribution from the lower bound in the thin interface limit $\xi/R \rightarrow 0$ is irrelevant.
The similarity of $\sigma_{\alpha \beta}^M$ with the usual domain contribution $\sigma_{\alpha \beta}^D$ (eq.~(\ref{sigma_D_domain})) is striking.  
The dimensionless coupling tensor
\begin{eqnarray}
s_{\alpha \beta} = \frac{4 \pi K_{\epsilon}E^{ex}_{\alpha}E^{ex}_{\beta} R}{\Gamma}
\label{s_coupling}
\end{eqnarray}
measures the relative importance of the electrostatic contribution to the interfacial tension at the scale of the domain size.
We see that the Maxwell stress contribution leads to an anisotropic renormalization of the interfacial tension.

{\it Comparison with experiment}---
We now attempt a quantitative comparison of our prediction with an experimental result.
Recently, Orihara et. al. have examined in detail the behaviors of an immiscible blend subjected to a step AC electric field ${\vec E}^{ex}=(0,0,E^{ex})$ under steady shear flow ${\vec v}_{flow}=({\dot \gamma}z, 0, 0)$~\cite{Orihara2011}.
In their experiment, two polymers with equal viscosity were blended, 
the minority phase of which (with the average volume fraction $\langle \phi_B \rangle = 1/9$) forms dispersed droplets in the absence of an electric field.
Upon the application of the step electric field, they observed the droplets elongation along the electric field, their coalescence, leading to the network structure formation. During this transient process, they measured the shear stress $\sigma_{xz}$, and at the same time, acquired the three-dimensional images with a confocal scanning laser microscope. This allowed them to quantitatively estimate the interfacial tensor $q_{\alpha \beta}$ and the area density $Q$. In addition, by intermittently turning off and on the electric field and the shear flow, they decomposed the total shear stress as $ \sigma_{xz}=\sigma^{V}_{xz}+\sigma^{D}_{xz}+\sigma^{M}_{xz}$. From these measurements, they confirmed the relation $\sigma^{V}_{xz}= \eta_0 {\dot \gamma}$ for the viscous stress,  $\sigma^{D}_{xz}=-\Gamma q_{xz}$ for the interfacial stress (eq.~(\ref{sigma_D_domain})) as expected. Furthermore, they found the proportionality relation between the electric component $\sigma^{M}_{xz}$ and 
 the interfacial tensor $q_{xz}$;
\begin{eqnarray}
\sigma^{M}_{xz} = -A^{(exp)}\frac{q_{xz}}{Q}
\label{sigmaM_q_exp}
\end{eqnarray}
with the slope $A^{(exp)}=64$ (Pa). 

Our formula eq.~(\ref{M_stress_expression}) applied to their experimental situation reads
\begin{eqnarray}
\sigma^{M}_{x z}  
\simeq -\Gamma q_{xz}s_{zz}
\simeq - 4 \pi K_{\epsilon}(E^{ex})^2 \langle \phi_B \rangle \frac{q_{xz}}{Q}
\end{eqnarray}
where we have used the relation $Q \simeq \langle \phi_B \rangle /R$.
Substituting the experimental parameters  $\langle \phi_B \rangle = 1/9$, $E^{ex}=6/\sqrt{2}$ kV/mm (effective value of the AC field),  $\epsilon_A = 2.7 \epsilon_0, \ \epsilon_B = 16 \epsilon_0$ (with $\epsilon_0$ being the vacuum permittivity)
\cite{Orihara2011}, 
we obtain the theoretical value $A^{(th)}=4 \pi K_{\epsilon}(E^{ex})^2  \langle \phi_B \rangle  = 83.7$ (Pa) in good agreement with the experimental value. 
This validates the present formula eq.~(\ref{M_stress_expression}) up to a numerical constant, which will enable us to investigate rheological response of domains in a systematic manner by solving the time-evolution equations for the interfacial tensor. 

{\it Summary}---
It is well known in electrostatics that a Maxwell stress is created at the boundary with dielectric gap.
It would be therefore natural to expect that the Maxwell stress in immiscible blends should be correlated with the interface configuration. 
We have demonstrated that this is indeed the case, and our formula (eq.~(\ref{M_stress_expression})) identifies the Maxwell stress contribution as an excess interfacial tension.
Importantly, this renormalization of the interfacial tension is anisotropic, i.e., active only for stress components connected with the external electric field through
the 
coupling tensor (eq.~(\ref{s_coupling})).

One may question the accuracy of the pre-averaging approximation (eq.~(\ref{preaverage})). Indeed, the pre-averaging was performed in an isotropic state, so we expect that it retains its physical justification when the magnitude of the coupling constant (eq.~(\ref{s_coupling})) is small.
Such a situation is realized when a strong shear flow is applied in which a factor controlling the domain size is the balance between surface and viscous stresses~\cite{Doi_Ohta91}, i.e., $R \simeq \Gamma/(\eta_0 {\dot \gamma})$, where ${\dot \gamma}$ is the shear rate.
Therefore, we find an alternative expression for the coupling tensor
\begin{eqnarray}
s_{\alpha \beta} \simeq \frac{4 \pi K_{\epsilon}E^{ex}_{\alpha}E^{ex}_{\beta} }{\eta_0 {\dot \gamma}}
\label{s_coupling_2}
\end{eqnarray}
This expression implies that the cross coupling between the flow and the electric field is highly nontrivial. 
The inverse of its trace $s_{\alpha \alpha}^{-1} \simeq \eta_0 {\dot \gamma}/[4 \pi K_{\epsilon}(E^{ex})^2]$ is known as a Mason number in the electro-rheological literature.

A remark is in order. The Maxwell stress plays a central role in the rheology of magneto-responsive fluids where magnetic colloids constitute anisotropic clusters such as a chain-like structure under magnetic field~\cite{Liu09}.
However, to our knowledge, there are no theories to connect the Maxwell stress with the interfacial tensor as has been formulated in the present paper.
%

Finally, although semi-quantitative agreement with the experiment is encouraging, we have to note that the formula~(\ref{M_stress_expression}) has been derived by assuming that $\delta \epsilon$ is small. With this limitation in mind, 
we expect that the proposed Maxwell stress formula in terms of the interfacial tensor provides a natural route to construct a coarse grained description of the electro-rheology of immiscible blends.

We thank H. Orihara and Y.H. Na for fruitful discussions, and I. Zaid for careful reading of the manuscript.
This work was supported by the JSPS Core-to-Core Program ``International research network for non-equilibrium dynamics of soft matter''.

\end{document}